\documentclass[onecolumn,10pt,showpacs,
amsmath,amssymb]{revtex4}
\usepackage{graphicx}
\usepackage{hyperref}
\usepackage{amssymb, amsmath}
\usepackage{epsf}

\def\la{\; \raise0.3ex\hbox{$<$\kern-0.75em\raise-1.1ex\hbox{$\sim$}}\;}
\def\ga{\;  \raise0.3ex\hbox{$>$\kern-0.75em\raise-1.1ex\hbox{$\sim$}}\;}

\def\pFn{p_{\raise-0.3ex\hbox{{\scriptsize F$\!$\raise-0.03ex\hbox{\rm n}}}}
}  
\def\pFp{p_{\raise-0.3ex\hbox{{\scriptsize F$\!$\raise-0.03ex\hbox{\rm p}}}}
}  
\def\pFe{p_{\raise-0.3ex\hbox{{\scriptsize F$\!$\raise-0.03ex\hbox{\rm e}}}}
}  
\def\pFmu{p_{\raise-0.3ex\hbox{{\scriptsize F$\!$\raise-0.03ex\hbox{
\rm $\mu$}}}} }  
\def\m@th{\mathsurround=0pt }
\def\eqalign#1{\null\,\vcenter{\openup1\jot \m@th
   \ialign{\strut$\displaystyle{##}$&$\displaystyle{{}##}$\hfil
   \crcr#1\crcr}}\,}
\def\dd{\mbox{d}}

\begin{document}

\title{Damping of sound waves in superfluid
nucleon-hyperon matter of neutron stars}
%
\author{Elena M. Kantor, Mikhail E. Gusakov}
\affiliation{
Ioffe Physical Technical Institute,
Politekhnicheskaya 26, 194021 Saint-Petersburg, Russia
}
\date{}
%

\pacs{
97.60.Jd,
26.60.+c,
47.37.+q,
47.75.+f        
}

\begin{abstract}
We consider sound waves
in superfluid nucleon-hyperon matter
of massive neutron-star cores.
We calculate and analyze the speeds
of sound modes
and their damping times
due to the shear viscosity and
non-equilibrium weak processes
of particle transformations.
For that, we employ the dissipative
relativistic hydrodynamics
of a superfluid nucleon-hyperon mixture,
formulated recently
(M. E. Gusakov and E. M. Kantor,
Phys. Rev. \textbf{D78}, 083006 (2008)).
We demonstrate that
the damping times of sound modes
calculated using this hydrodynamics
and the ordinary (nonsuperfluid) one,
can differ from each other
by several orders of magnitude.
\end{abstract}

\maketitle

\section{Introduction}
\label{1}

In recent years there is a growing interest
in studies of neutron-star pulsations.
This is related to a number of reasons.
First of all, the recently discovered 
high frequency oscillations of
electromagnetic radiation during giant flares 
may be associated 
with the pulsations of neutron stars 
\cite{sw2005,israel2005}.
Second, the gravitational-wave detectors, 
which will be able
to detect gravitational radiation 
from isolated pulsating neutron stars,
are under construction 
\cite{ak01,andersson03,andersson06,ligo07}.

For interpretation of the observations, 
it is important to have a well-developed theory of 
neutron-star pulsations.
The construction of such a theory is complicated
by the fact that the baryons in neutron-star cores
can be in superfluid state \cite{yls99,ls01,yp04,bb98, vt04, tnyt06}.
Thus, to study the pulsations, 
one has to use a hydrodynamics 
describing mixtures of superfluid liquids.
There is a substantial body of literature,
devoted to pulsations of superfluid neutron stars
(see, e.g., Refs.\ \cite{epstein88,ac01,ga06,gusakov07,
lm94,lee95,ac01a,acl02,pca04,yl03a,yl03b,
cll99,sac08,lac08}).
All these papers deal 
with the nucleon $npe(\mu)$ matter 
composed of neutrons ($n$), protons ($p$), 
and electrons ($e$) 
with possible admixture of muons ($\mu$).
Most of them study pulsations at zero temperature 
(see, however, Refs.\ \cite{ga06,gusakov07}). 

In this paper we for the first time 
investigate dynamic properties of
superfluid {\it nucleon-hyperon} matter
in the cores of massive neutron stars,
composed,
in addition to neutrons, protons, electrons, 
and muons,
of $\Lambda$ and $\Sigma^{-}$
hyperons ($\Lambda$ and $\Sigma$, respectively).
For that, we employ the relativistic 
hydrodynamics \cite{gk08},
describing a superfluid nucleon-hyperon mixture 
at {\it arbitrary} temperature.
We study the simplest pulsations
in such matter -- sound modes,
how they travel and how they damp.
Within this simple example we demonstrate, 
in particular, 
that the characteristic damping times of pulsations,
calculated self-consistently 
in the frame of superfluid hydrodynamics,
can differ by several orders of magnitude from those
calculated 
using the
nonsuperfluid hydrodynamics
(in the latter case 
the effects of superfluidity are taken into account 
only at calculating kinetic coefficients).

The paper is organized as follows.
In Sec.\ II we give an overview of the main reactions
of particle mutual transformations
in the nucleon-hyperon matter.
In Sec.\ III we briefly discuss the 
relativistic dissipative hydrodynamics describing 
superfluid nucleon-hyperon mixture.
In Sec.\ IV we analyze the sound modes in such matter 
neglecting dissipation.
In Sec.\ V we calculate the damping times of sound modes
due to the nonequilibrium reactions 
(\ref{s})--(\ref{ll}) (see Sec. II)
and shear viscosity.
Section VI presents summary.

Throughout the paper,
unless otherwise stated, 
we use the system of units
in which the Plank constant $\hbar$,
the speed of light $c$
and the Boltzmann constant $k_{\rm B}$
equal unity, 
$\hbar=c=k_{\rm B}=1$.

\section{The main processes of particle
transformations in nucleon-hyperon matter}
\label{11}

The most effective weak processes in nucleon-hyperon matter
are the following nonleptonic reactions
\cite{jones01,hly02,lo02,schaffner08}
\begin{eqnarray}
n+n &\leftrightarrow& p + \Sigma^-,
\label{s}\\
n+p &\leftrightarrow& p +\Lambda,
\label{l}\\
n+n &\leftrightarrow& n+ \Lambda,
\label{ln}\\
n+\Lambda &\leftrightarrow& \Lambda+ \Lambda.
\label{ll}
\end{eqnarray}
The full thermodynamic equilibrium
implies, in particular, the equilibrium with respect
to these reactions,
\begin{eqnarray}
\delta \mu_{1}\equiv2 \mu_{n0} - \mu_{p 0} - \mu_{\Sigma 0} =0,
\label{s_eq} \\
\delta \mu_{2}=\delta \mu_{3}=\delta \mu_{4}\equiv \mu_{n0}-\mu_{\Lambda 0}=0.
\label{l_eq}
\end{eqnarray}
In this case the average number 
of direct reactions in unit volume per unit time 
is equal to the number 
of inverse reactions.
In Eqs.\ (\ref{s_eq}) and (\ref{l_eq})
$\mu_{i0}$ are the chemical potentials
of particle species $i=n$, $p$, $\Lambda$, and $\Sigma$,
taken at equilibrium;
$\delta \mu_{m}$ ($m=1$,$\ldots$, $4$) are the disbalances
of the chemical potentials for the reactions
(\ref{s})--(\ref{ll}).
In what follows, we mark the equilibrium values 
of thermodynamic quantities with the subscript $0$.
Accordingly, the thermodynamic quantities without the subscript $0$
(e.g., $\mu_{i}$) refer to perturbed matter.
Notice that, the equilibrium conditions for the reactions
(\ref{l}), (\ref{ln}), and (\ref{ll}) coincide.

Eqs.\ (\ref{s_eq}) and (\ref{l_eq}) do not hold
in the perturbed matter
($\delta \mu_1\neq 0$,
$\delta \mu_2=\delta \mu_3=\delta \mu_4\neq 0$),
so that the numbers of direct and inverse reactions 
are not equal.
The nonequilibrium reactions 
(\ref{s})--(\ref{ll}) act to return
the system to the equilibrium state.
This leads to dissipation of mechanical energy, 
accumulated in the matter.

Along with the reactions (\ref{s})--(\ref{ll}) 
there is a number of weak reactions with leptons. 
The leptonic reactions
(e.g, the direct and modified Urca processes
with electrons or muons)
are much slower in comparison to
the reactions (\ref{s})--(\ref{ll}).
In the interesting range of parameters 
(temperatures and pulsation frequencies),
they cannot influence substantially
chemical composition of the perturbed matter
and will be neglected in what follows.
However, we assume that the unperturbed matter satisfies the
equilibrium conditions with respect to these reactions,
\begin{equation}
\mu_{n0}=\mu_{p0}+\mu_{e0}, \qquad \mu_{e0}=\mu_{\mu 0},
\label{lepton_cond}
\end{equation}
where $\mu_{e0}$ and $\mu_{\mu0}$ 
are the equilibrium 
chemical potentials for electrons and muons, 
respectively.
In addition to the processes described above,
there is a fast nonleptonic reaction
due to the strong interaction of baryons,
\begin{equation}
n+\Lambda \leftrightarrow p + \Sigma^-.
\label{fast}
\end{equation}
We assume that the perturbed matter is always
in equilibrium with respect to this reaction \cite{lo02,gk08},
\begin{equation}
\delta \mu_{\rm fast} \equiv
\mu_{n}+\mu_\Lambda-\mu_{p}-\mu_\Sigma =0.
\label{eqfast}
\end{equation}
It follows from the condition (\ref{eqfast}),
that 
the chemical potential disbalances
for all the four reactions (\ref{s})--(\ref{ll}) coincide,
%
\begin{equation}
\delta \mu \equiv  \delta \mu_1
= \delta \mu_2 =\delta \mu_3=\delta \mu_4.
\label{def}
\end{equation}
Now the equilibrium conditions (\ref{s_eq}) and (\ref{l_eq})
can be rewritten as
\begin{equation}
\delta \mu=0.
\label{dmu1}
\end{equation}
Below we denote the difference between the average number 
of direct and inverse reactions
(\ref{s}),$\ldots$,(\ref{ll}),
occurring in the unit volume per unit time,
by $\Delta \Gamma_1$,$\ldots$,$\Delta \Gamma_4$, 
respectively.
In this paper we consider
small deviations from the
chemical equilibrium, 
$\delta \mu \ll T$.
In this case the quantities $\Delta \Gamma_m$ ($m=1$,$\ldots$,4)
can be expanded in powers of $\delta \mu$
and presented in the linear approximation as
(see, e.g., \cite{hly02,lo02,gk08}):	
\begin{equation}
\Delta \Gamma_m=\lambda_m \delta \mu,
\label{source}
\end{equation}
where $\lambda_m$ are the rates of the reactions
(\ref{s})--(\ref{ll}),
some functions of the number densities and temperature.

\section{The relativistic hydrodynamics
of a superfluid nucleon-hyperon mixture}
\label{2}
In this and subsequent sections, 
the subscripts $i$ and $k$ refer
to baryons ($i,k = n, p, \Lambda, \Sigma$).
The summation is assumed over
repeated baryon indices $i$~and~$k$.
The subscript $l$ refers to leptons
($l=e, \mu$); the subscript $j$
runs over all particle species
($j=n$, $p$, $\Lambda$, $\Sigma$, $e$, $\mu$);
$\mu$ and $\nu$ are the space-time indices.

The relativistic hydrodynamics, 
describing a nucleon-hyperon mixture,
composed of superfluid neutrons,
protons, $\Lambda$ and $\Sigma^-$ hyperons,
as well as normal electrons and muons,
has been formulated in Ref.\ \cite{gk08}.
It is a direct generalization
of the hydrodynamics of superfluid $npe$ matter 
\cite{gusakov07}.
In this section 
we briefly discuss
the main equations of this hydrodynamics. 

\subsection{The nondissipative hydrodynamics}
\label{2a}

Let us discuss first the nondissipative hydrodynamics
of superfluid nucleon-hyperon matter.
Equations of superfluid hydrodynamics include
the continuity equations for particle species~$j$
\begin{equation}
\partial_{\mu} j^{\mu}_{ (j) } = 0,
\label{particle_conservation}
\end{equation}
where the particle 
four currents $j^{\mu}_{ (j) }$ equal
\begin{equation}
j^{\mu}_{(i)} = n_i u^{\mu} + Y_{ik} w^{\mu}_{(k)}, \quad
j^{\mu}_{({l})} = n_{l} u^{\mu};
\label{particle_conservation1}
\end{equation}
energy-momentum conservation law
\begin{equation}
\partial_{\mu} T^{\mu \nu} =0,
\label{energymomentum}
\end{equation}
where the energy-momentum tensor $T^{\mu \nu}$ equals
\begin{eqnarray}
&& T^{\mu \nu} = (P+\varepsilon) \, u^{\mu} u^{\nu}
+ P \eta^{\mu \nu}
\nonumber \\
&&+ Y_{ik} \left[ w^{\mu}_{(i)} w^{\nu}_{(k)}
+ \mu_i \, w^{\mu}_{(k)} u^{\nu}
+ \mu_k \, w^{\nu}_{(i)} u^{\mu} \right];
\label{Tmunu2}
\end{eqnarray}
the second law of thermodynamics
\begin{equation}
\dd \varepsilon = T \, \dd S +
\mu_i \, \dd n_i + \mu_e \, \dd n_e
+\mu_{\mu} \, \dd n_{\mu}+ { Y_{ik} \over 2}
\,
\dd \left[ w^{\mu}_{(i)} w_{(k) \mu} \right],
\label{2ndlaw2}
\end{equation}
%
and a number of conditions for superfluid components,
which are specified below
(see Eqs.\ (\ref{uw}) and (\ref{w_i2})).
In Eqs.\ (\ref{particle_conservation})--(\ref{2ndlaw2})
$\eta^{\mu \nu}={\rm diag}(-1,+1,+1,+1)$ 
is the special relativistic metric;
$n_j$ is the number density 
of particle species $j$;
$\varepsilon$, $T$, and $S$ are the energy density, 
temperature, and entropy density, respectively;
$P$ is the pressure, which is defined by the same formula
as for nonsuperfluid matter \cite{ga06,gusakov07}
\begin{equation}
P=-\varepsilon + \mu_i n_i + \mu_e n_e + \mu_{\mu} n_{\mu} + T S;
\label{Pressure}
\end{equation}
$u^{\mu}$ is the four velocity of the normal (nonsuperfluid)
liquid component, normalized so that
\begin{equation}
u_{\mu} u^{\mu}=-1.
\label{normirovka}
\end{equation}
(we assume that all the nonsuperfluid components move
with the same velocity $u^{\mu}$).
Next, $Y_{ik}$ is the symmetric
relativistic entrainment matrix
of nucleon-hyperon mixture,
calculated in Ref.\ \cite{gkh08a} 
for zero temperature and in Ref.\ \cite{gkh08b}
for arbitrary temperature.
The important property is that 
if some particles (e.g., neutrons) are nonsuperfluid
then the related elements of this matrix vanish, 
$Y_{nk}=Y_{kn}=0$ ($k=n, p, \Lambda, \Sigma$).
In the nonrelativistic limit 
this 
$4 \times 4$ matrix
is expressed through the nonrelativistic 
entrainment matrix $\rho_{ik}$
as \cite{ga06,gusakov07}:
$Y_{ik}=\rho_{ik}/(m_i m_k)$,
where $m_i$ is the mass of a free baryon species $i$
(the matrix $\rho_{ik}$ is a generalization
of the superfluid density to the case of mixtures,
see, e.g., Refs.\ \cite{ab75,bjk96,gh05}).
The motion of superfluid component of a species $i$
is described by the four vector $w^{\mu}_{(i)}$, 
which meets the condition
\begin{equation}
u_{\mu} w^{\mu}_{(i)}=0.
\label{uw}
\end{equation}
The potentiality of superfluid motion
is expressed as
\begin{eqnarray}
&&\partial^{\nu} \left[ w^{\mu}_{(i)}
+q_i A^{\mu} + \mu_i u^{\mu} \right]
\nonumber \\
&&= \partial^{\mu} \left[ w^{\nu}_{(i)}
 +q_i A^{\nu} +\mu_i u^{\nu} \right],
\label{w_i2}
\end{eqnarray}
where $q_i$ is the electric charge of particle species $i$;
$A^\mu$ is the four potential of the electromagnetic field.
The potentiality condition (\ref{w_i2}) 
is equivalent to a statement
that there is a scalar function $\phi_i$, 
satisfying (see Ref.\ \cite{gusakov07,gk08})
\begin{equation}
\partial^{\mu} \phi_i= w^{\mu}_{(i)}
+ q_i A^{\mu} + \mu_i u^{\mu}.
\label{w_i}
\end{equation}
The scalar $\phi_i$ is related to the wave function phase of the
Cooper-pair condensate $\Phi_i$ by the equality
${\pmb \triangledown} \phi_i={\pmb \triangledown\Phi_i}/2$.
In the nonrelativistic limit the spatial parts
${\pmb u}$ and ${\pmb w}_{(i)}$
of the four vectors $u^\mu$ and
$w^\mu_{(i)}$ transform into
\begin{eqnarray}
{\pmb u}={\pmb V}_q, \quad \quad
{\pmb w}_{(i)}=m_i({\pmb V}_{s(i)}-{\pmb V}_q),
\end{eqnarray}
where ${\pmb V}_q$ and
${\pmb V}_{s(i)}=({\pmb \nabla} \phi_i-q_i {\pmb A})/m_i$
are, respectively, 
the normal and superfluid velocities
of the nonrelativistic theory
of superfluid liquids \cite{khalatnikov89,putterman74}.

The hydrodynamics described above would be incomplete
without an indication in what reference frame we define (measure)
the main thermodynamic quantities 
(i.e., what frame is comoving).
As was demonstrated in Ref.\ \cite{gusakov07},
the condition (\ref{uw}) dictates that
the comoving is the frame where
the four velocity $u^{\mu}$ equals $u^{\mu}=(1,0,0,0)$.
In this frame, the basic thermodynamic quantities
$\varepsilon$, $n_j$, and ${\pmb w}_{(i)}$
(or ${\pmb \triangledown \phi_i}$) are defined by
(see Eqs.\ (\ref{particle_conservation1}),
(\ref{Tmunu2}), and (\ref{uw}))
\begin{eqnarray}
j_{j}^0 &=& n_j, \quad \quad
\label{condition1} \\
{\pmb j}_{i} &=& Y_{ik} \, {\pmb w}_{(k)}
= Y_{ik} \, \left( {\pmb \triangledown} \phi_k -q_k {\pmb A} \right), \quad
\quad
\label{condition2} \\
T^{00} &=& \varepsilon.
\label{condition3}
\end{eqnarray}
All other thermodynamic quantities 
in nonequilibrium matter
are the same functions of 
$\varepsilon$, $n_j$, and ${\pmb w}_{(i)}$
(or, equivalently, 
$\varepsilon$, $n_j$, and $w^{\mu}_{(i)}w_{(k) \mu}$)
as in the full thermodynamic equilibrium.

\subsection{Viscous dissipation}
\label{2b}

The main
dissipative mechanisms
in the pulsating nucleon-hyperon matter are
the shear viscosity and effective bulk viscosity 
due to the nonequilibrium processes (\ref{s})--(\ref{ll}).
These are the two mechanisms which 
will be analyzed in this paper.

The shear viscosity leads to an additional term
in the energy-momentum tensor $T^{\mu \nu}$.
It now takes the form
\begin{eqnarray}
T^{\mu \nu} &=& (P+\varepsilon) \, u^{\mu} u^{\nu}
+ P \eta^{\mu \nu}
\nonumber \\
&+& Y_{ik} \left[ w^{\mu}_{(i)} w^{\nu}_{(k)}
+ \mu_i \, w^{\mu}_{(k)} u^{\nu}
+ \mu_k \, w^{\nu}_{(i)} u^{\mu} \right]
+ \tau^{\mu \nu}_{\rm sh},
\label{Tmunu3}
\end{eqnarray}
where
\begin{equation}
\tau^{\mu \nu}_{\rm sh}=
- \eta \, H^{\mu \gamma} \, H^{\nu \delta} \,\,
\left( \partial_{\delta} u_{\gamma} + \partial_{\gamma} u_{\delta}
- {2 \over 3} \,\, \eta_{\gamma \delta} \,\,
\partial_{\varepsilon} u^{\varepsilon}  \right);
\label{taush1}
\end{equation}
$H^{\mu \nu} \equiv \eta^{\mu \nu} + u^{\mu} u^{\nu}$
is the projection matrix;
$\eta$ is the shear viscosity coefficient.

As has been already mentioned, 
the effect of the nonequilibrium processes 
(\ref{s})--(\ref{ll}) can be described 
in terms of the bulk viscosity formalism
(for more details, see \cite{hly02,lo02,gk08}).
However, for the analysis of sound modes 
it is more convenient
to take these processes into account explicitly,
by introducing sources 
into the right-hand sides 
of the continuity equations (\ref{particle_conservation}),
\begin{equation}
\partial_{\mu} j^{\mu}_{ (j) } = \Delta S_j,
\label{particle_conservation2}
\end{equation}
similar to how it was done for normal (nonsuperfluid) 
$npe$ matter in Ref.\ \cite{gyg05}.
Here $\Delta S_j$ is a number 
of particles of a species $j$,
generated in the unit volume per unit time
in the reactions (\ref{s})--(\ref{ll}).
Since we neglect the leptonic reactions,
$\Delta S_e=\Delta S_{\mu}=0$.

As a result, the hydrodynamics describing 
a {\it viscous} superfluid 
nucleon-hyperon mixture differs from
the nondissipative hydrodynamics of Sec.\ IIIA only
by the expression for the energy-momentum tensor
(Eq.\ (\ref{Tmunu3}) instead of (\ref{Tmunu2}))
and by the continuity equations
(Eq.\ (\ref{particle_conservation2})
instead of (\ref{particle_conservation})).
Using this hydrodynamics, one can derive
the entropy generation equation
(see also \cite{gusakov07,gk08})
\begin{equation}
T \, \partial_{\mu} \left(S u^{\mu} \right) =
\delta \mu \, \Delta \Gamma
- \tau^{\mu \nu}_{\rm sh} \,\, \partial_{\mu} u_{\nu},
\label{entropy2}
\end{equation}
where we define
\begin{equation}
\Delta \Gamma \equiv \Delta \Gamma_1 + \Delta \Gamma_2
+\Delta \Gamma_3+\Delta \Gamma_4 = \lambda \, \delta \mu,
\label{dG}
\end{equation}
with $\lambda \equiv \lambda_1+\lambda_2+\lambda_3+\lambda_4$
(see Eq.\ (\ref{source})).

\section{Sound waves in the absence of dissipation}
\label{3}


\subsection{The main assumptions}
\label{3a}

We assume that
all the hydrodynamic velocities 
equal zero in the equilibrium,
$u^{\mu}=(1,0,0,0)$ and $w^{\mu}_{(i)}=(0,0,0,0)$
($i=n$, $p$, $\Lambda$, $\Sigma$).
(In principle, 
it is not necessary 
to have vanished all the components of the 
four vector $w^{\mu}_{(i)}$.
It is well known, 
that even in thermodynamic equilibrium
a motion is possible with 
nonzero superfluid velocities \cite{khalatnikov89}.
This means that generally, 
the spatial components of the four vector
$w^{\mu}_{(i)}$ can be nonzero.
As for the time component $w^0_{(i)}$,
it vanishes in any case,
as it follows from Eq.\ (\ref{uw}).)

We assume also
that in the perturbed matter
the quasineutrality condition holds
\begin{equation}
n_p = n_e + n_{\mu} + n_{\Sigma},
\label{charge neutrality}
\end{equation}
in addition
to the equilibrium condition (\ref{eqfast})
with respect to the fast reaction (\ref{fast}).
Using Eq.\ (\ref{charge neutrality})
and the continuity equations (\ref{particle_conservation2}) 
for protons, $\Sigma^-$-hyperons, electrons and muons, 
as well as the fact that 
$\Delta S_p-\Delta S_e -\Delta S_{\mu}-\Delta S_{\Sigma}=0$,
one obtains \cite{gk08}
\begin{equation}
\partial_{\mu} \left[ Y_{pk} w^{\mu}_{(k)} \right]
= \partial_{\mu} \left[Y_{\Sigma k} w^{\mu}_{(k)} \right].
\label{quasicond}
\end{equation}
In this paper we consider small deviations 
from the equilibrium state.
Thus, we restrict ourselves to linear terms 
in perturbation. 
As follows from 
the normalization condition 
(\ref{normirovka}) and Eq.\ (\ref{uw}), 
in the linear approximation
the time components $u^0$ and $w^0_{(i)}$
in the perturbed matter
remain the same,
\begin{equation}
u^0=1, \qquad
w^0_{(i)}=0,
\label{time_components}
\end{equation}
while their derivatives vanish
\begin{equation}
\partial_{\mu} u^0 =0, \qquad
\partial_{\mu} w^0_{(i)}=0.
\label{derivatives}
\end{equation}
Using Eqs.\ (\ref{time_components}) 
and (\ref{derivatives}),
one gets from Eq.\ (\ref{quasicond})
\begin{equation}
{\rm div} \left[ (Y_{pk}-Y_{\Sigma k}) {\pmb w}_{(k)} \right]=0.
\label{sound5g}
\end{equation}

As we have already emphasized in Sec.\ IIIA,
in the nonequilibrium matter
any thermodynamic quantity
(e.g., $P$ or $\mu_j$)
is a function of the number densities $n_j$, 
temperature $T$, 
and scalars $w^{\mu}_{(i)}w_{(k) \mu}$
(we choose $T$ instead of $\varepsilon$ 
as an independent variable).
In the linear approximation we can neglect the dependence
on the quadratically small quantity $w^{\mu}_{(i)}w_{(k) \mu}$.
Moreover, in the strongly degenerate 
nucleon-hyperon matter,
the dependence of $P$ and $\mu_j$ on $T$
can also be neglected
(see, e.g., \cite{reisenegger95,gyg05,gusakov07}).
Consequently, $P$ and $\mu_j$ 
are functions of only $n_j$
($j=n$, $p$, $\Lambda$, $\Sigma$, $e$, $\mu$).
These six number densities 
are related by the conditions
(\ref{eqfast}) and (\ref{charge neutrality}),
so that only $6-2=4$ of them are independent.
Thus, the pressure $P$ and the chemical potentials $\mu_j$
depend on some four number densities (or their functions).
As the independent variables it is convenient to choose 
the number density of baryons 
$n_b=n_n+n_p+n_{\Lambda}+n_{\Sigma}$,
the number density of hyperons
$n_H=n_{\Lambda}+n_{\Sigma}$, 
and the quantities 
$n_{\Sigma n}=n_n+n_{\Sigma}$
and $y=n_e/n_{\mu}$ (see Ref.\ \cite{gk08}).
Notice that, in the thermodynamic equilibrium
$n_b$, $n_H$, $n_{\Sigma n}$, and $y$
are additionally constrained by the two conditions
(\ref{lepton_cond}) and by the condition (\ref{dmu1}).
In this case $P$ and $\mu_j$
are functions of only one number density
(e.g., $n_b$).

\subsection{Equations governing the sound waves}
\label{3b}

In this section we derive the system of equations
describing sound waves
in the superfluid nucleon-hyperon matter
neglecting dissipation 
due to
the nonequilibrium reactions (\ref{s})--(\ref{ll})
and shear viscosity.

As we demonstrate below, 
the nonequilibrium reactions
do not lead to dissipation in two limiting cases
($i$) either in the limit of {\it slow} reactions,
when the total rate $\lambda$ 
of the reactions (\ref{s})--(\ref{ll})
is negligible, 
so that they cannot change the matter composition
during the pulsations excited by a sound wave
(i.e. $\Delta \Gamma_m =0$, $m=1$,$\ldots$,$4$);
($ii$) or in the limit of {\it fast} reactions,
when the reaction rate is so high
that the pulsating matter 
is always in equilibrium
with respect to the reactions (\ref{s})--(\ref{ll}),
so that the condition (\ref{dmu1}) is always satisfied.
These are two cases which will be analyzed in what follows.

The system of linearized hydrodynamic equations,
describing sound waves, 
consists of ($i$) the condition (\ref{sound5g});
($ii$) momentum conservation law
\begin{equation}
\partial_t \left[(P_0 + \varepsilon_0) \, {\pmb u}
+ \mu_{i0} Y_{ik} \, {\pmb w}_{(k)} \right]
= -{\pmb \triangledown}  P,
\label{sound1 0}
\end{equation}
following from Eq.\ (\ref{energymomentum});
and ($iii$) the four potentiality conditions 
for superfluid motion (for each baryon species)
\begin{eqnarray}
&&\partial_t
\left[
\mu_{n0} \, {\pmb u} + {\pmb w}_{(n)}
\right]
= - {\pmb \triangledown}  \mu_{n},
\label{sound2 0} \\
&&\partial_t
\left[
\mu_{\Lambda 0} \, {\pmb u} + {\pmb w}_{(\Lambda)}
\right]
= - {\pmb \triangledown}  \mu_{\Lambda},
\label{sound3 0} \\
&&\partial_t
\left[
\mu_{p 0} \, {\pmb u} + {\pmb w}_{(p)}+q_{p} {\pmb A}
\right]
= - {\pmb \triangledown} \left[ \mu_{p}+q_{p}  A^0 \right],
\label{sound4 0} \\
&&\partial_t
\left[
\mu_{\Sigma 0} \, {\pmb u} + {\pmb w}_{(\Sigma)}+q_{\Sigma} {\pmb A}
\right]
= - {\pmb \triangledown} \left[ \mu_{\Sigma}+q_{\Sigma}  A^0 \right].
\label{sound5 0}
\end{eqnarray}
These conditions can be obtained from Eq.\ (\ref{w_i2})
with the help of Eq.\ (\ref{derivatives}).
The vector potential ${\pmb A}$ and the scalar potential $A^0$
can be excluded 
from Eqs.\ (\ref{sound4 0}) and (\ref{sound5 0})
if one notices that $q_{\Sigma}=-q_p$
and takes a sum of these formulas.
Subtracting then Eqs.\ (\ref{sound2 0}) and (\ref{sound3 0})
from the obtained sum and using 
the equilibrium condition (\ref{eqfast}), 
one gets
\begin{equation}
\partial_t \left[{\pmb w}_{(\Sigma)}+{\pmb w}_{(p)}
-{\pmb w}_{(\Lambda)}-{\pmb w}_{(n)}
\right] =0.
\label{sound4}
\end{equation}

In view of the condition (\ref{l_eq}),
and the definition (\ref{def}),
it follows from Eqs.\ (\ref{sound2 0}) 
and (\ref{sound3 0}) that
\begin{equation}
\partial_t
\left[
{\pmb w}_{(n)} - {\pmb w}_{(\Lambda)}
\right]
= - {\pmb \triangledown}  \delta \mu.
\label{sound5}
\end{equation}

Eqs.\ (\ref{sound5g})--(\ref{sound2 0}),
(\ref{sound4}), and (\ref{sound5})
represent the five equations for five variables,
${\pmb u}$ and ${\pmb w}_{(i)}$
($i=n$, $p$, $\Lambda$, $\Sigma$).
They completely describe the sound waves
in the superfluid nucleon-hyperon matter 
under the condition that
the pressure $P(n_b, n_H, n_{\Sigma n}, y)$,
neutron chemical potential
$\mu_n(n_b, n_H, n_{\Sigma n}, y)$,
and chemical potential disbalance
$\delta \mu(n_b, n_H, n_{\Sigma n}, y)$
are known as functions of 
${\pmb u}$ and ${\pmb w}_{(i)}$.
Expanding these quantities in Taylor series
near the equilibrium point and presenting $P$ and $\mu_n$
as $P=P_0 + \delta P$ and
$\mu_n = \mu_{n0} + \delta \mu_{n}$,
one obtains
\begin{eqnarray}
\delta P =  \frac{\partial P}{\partial n_{b}} \, \delta n_b
+ \frac{\partial P}{\partial n_{H}} \, \delta n_H
+\frac{\partial P}{\partial n_{\Sigma n}} \, \delta n_{\Sigma n}
+\frac{\partial P}{\partial y} \, \delta y,
\label{delta P}\\
\delta \mu_{\rm n} =  \frac{\partial \mu_{\rm n}}{\partial n_{b}} \, \delta n_b
+ \frac{\partial \mu_{\rm n}}{\partial n_{H}} \, \delta n_H
+\frac{\partial \mu_{\rm n}}{\partial n_{\Sigma n}} \, \delta n_{\Sigma n}
+\frac{\partial \mu_{\rm n}}{\partial y} \, \delta y,
\label{delta mun}\\
\delta \mu=  \frac{\partial \delta \mu}{\partial n_{b}} \,
\delta n_b
+ \frac{\partial \delta \mu}{\partial n_{H}} \, \delta n_H
+\frac{\partial \delta \mu}{\partial n_{\Sigma n}} \, \delta n_{\Sigma n}
+\frac{\partial\delta \mu}{\partial y} \, \delta y,
\label{dmu}
\end{eqnarray}
where in the last equation
we take into account that $\delta \mu=0$ 
in the equilibrium state (see Eq.\ (\ref{dmu1})).
In Eqs.\ (\ref{delta P})--(\ref{dmu})
$\delta n_b$, $\delta n_H$, $\delta n_{\Sigma n}$, and $\delta y$
are the deviations of the quantities $n_b$, $n_H$, $n_{\Sigma n}$, and $y$
from their equilibrium values
$n_{b0}$, $n_{H0}$, $n_{\Sigma n0}$, and $y_0$, respectively.
In the Appendix these deviations are expressed 
through the velocities
${\pmb u}$ and ${\pmb w}_{(i)}$ 
in the limit of {\it slow}
and {\it fast} reactions.

Assuming now, 
that the perturbations are harmonic
($\sim {\rm e}^{i(\omega t + {\pmb k} {\pmb r})}$,
where $\omega$ is the pulsation frequency
and ${\pmb k}$ is the wave vector),
the system of Eqs.\ (\ref{sound5g})--(\ref{sound2 0}),
(\ref{sound4}), and (\ref{sound5})
can be rewritten as
\begin{eqnarray}
&&  (Y_{p k}-Y_{{\Sigma} k}) \, {\pmb w}_{(k)}=0,
\label{sound55g} \\
&& i \omega
\left[
(P_0 + \varepsilon_0) \, {\pmb u}
+ \mu_{i0} Y_{{i}k} \, {\pmb w}_{(k)}
\right]
= -i {\pmb k}  \, \delta P,
\label{sound1g} \\
&&i \omega
\left[
\mu_{n0} \, {\pmb u} + {\pmb w}_{(n)}
\right]
= -i {\pmb k} \,  \delta  \mu_{n},
\label{sound2g} \\
&& {\pmb w}_{(\Lambda)}+{\pmb w}_{(n)}={\pmb w}_{(\Sigma)}+{\pmb w}_{(p)},
\label{sound4g} \\
&&i \omega
\left[
{\pmb w}_{(n)}-{\pmb w}_{(\Lambda)}
\right]
= -i {\pmb k} \,  \delta\mu.
\label{sound3g}
\end{eqnarray}
In the limit of {\it fast} reactions,
when the condition (\ref{dmu1}) holds,
Eq.\ (\ref{sound3g}) can be further simplified,
\begin{equation}
{\pmb w}_{(n)} - {\pmb w}_{(\Lambda)} = 0.
\label{sound33g}
\end{equation}

Eqs.\ (\ref{sound55g})--(\ref{sound3g})
together with (\ref{delta P})--(\ref{dmu}),
as well as the expressions 
(\ref{dnb slow})--(\ref{dnSn slow})
for the case of {\it slow} reactions
and (\ref{dnb fast})--(\ref{dnSn fast})
for the case of {\it fast} reactions,
allow one to determine the speeds of sound modes.

\subsection{Results for sound speeds}
\label{3c}

From the analysis of Eqs.\ (\ref{sound55g})--(\ref{sound3g})
it is clear, that the vectors $\pmb u$ and ${\pmb w_{(i)}}$
must be collinear with the wave vector ${\pmb k}$:
$\pmb u, \pmb w_{(i)} \parallel \pmb k$.
Thus, the system of equations
(\ref{sound55g})--(\ref{sound3g})
has the form
\begin{equation}
{\pmb A} \cdot {\pmb x}=0.
\label{Ax}
\end{equation}
Here ${\pmb x}$ is a vector that equals
${\pmb x} = (u,w_{(n)},w_{ (p)},
w_{(\Lambda)},w_{(\Sigma)})$,
where $u \equiv \pmb u \pmb k/k$
and $w_{(i)}\equiv \pmb w_{(i)} \pmb k/k$;
${\pmb A}$ is a $5 \times 5$ matrix,
which elements depend on 
the thermodynamic quantities (and their derivatives),
on the relativistic entrainment matrix $Y_{ik}$,
and on the frequency $\omega$ and the wave number $k$.

The system of equations (\ref{Ax}) 
has a nontrivial solution if
${\rm det} \, {\pmb A}=0$.
This condition results in a cubic equation
on the speed of sound $s \equiv \omega/k$ squared.
Each of three solutions to this equation
describes two sound waves, 
propagating with the same speed
along and opposite to the wave vector ${\pmb k}$.
In principle, 
these solutions can be written out analytically,
but here we do not present them 
because they are too lengthy.

In the case of {\it slow} reactions
(when the condition
$\lambda (\partial \delta \mu/\partial n_H
-\partial \delta \mu
/\partial n_{\Sigma n})\ll \omega$ holds, 
see Appendix), 
the three nonzero solutions to the cubic equation exist,
corresponding to three different sound modes.
In the case of {\it fast} reactions
(when $\lambda (\partial \delta \mu/\partial n_H
-\partial \delta \mu/\partial n_{\Sigma n}) \gg \omega$),
one of the three roots of the cubic equation vanishes,
hence the actual number of sound modes is two
(this conclusion can be also drawn from the fact that
in the limit of {\it fast} reactions
one can use Eq.\ (\ref{sound33g}) instead of (\ref{sound3g}),
which does not depend on ${\pmb k}$ and $\omega$).

The number of independent sound modes 
in these two limiting cases
can be easily understood 
from the following reasoning.
When all baryon species are superfluid,
there are 5 velocity fields in the matter, 
${\pmb u}$ and ${\pmb w}_{(i)}$.
In the limit of {\it slow} reactions 
the velocities are related by 
Eqs.\ (\ref{sound55g}) and (\ref{sound4g})
(these equations follow from
the quasineutrality condition (\ref{charge neutrality})
and the condition of the equilibrium
with respect to the fast reaction (\ref{eqfast}), 
respectively).
Thus, in this limit 
we have three independent velocities ($5-2=3$).
Correspondingly, the number of independent sound modes is also three.
In the limit of {\it fast} reactions,
there is a constraint (\ref{sound33g})
in addition to
the conditions (\ref{sound55g}) and (\ref{sound4g}).
This constraint is a consequence 
of the equilibrium condition (\ref{dmu1})
with respect to the reactions (\ref{s})--(\ref{ll}).
Thus, the number of independent velocities 
(and sound modes) equals two ($5-3=2$).

\begin{figure}
\begin{center}
\leavevmode \epsfxsize=9cm \epsfbox[100 300 400 500]{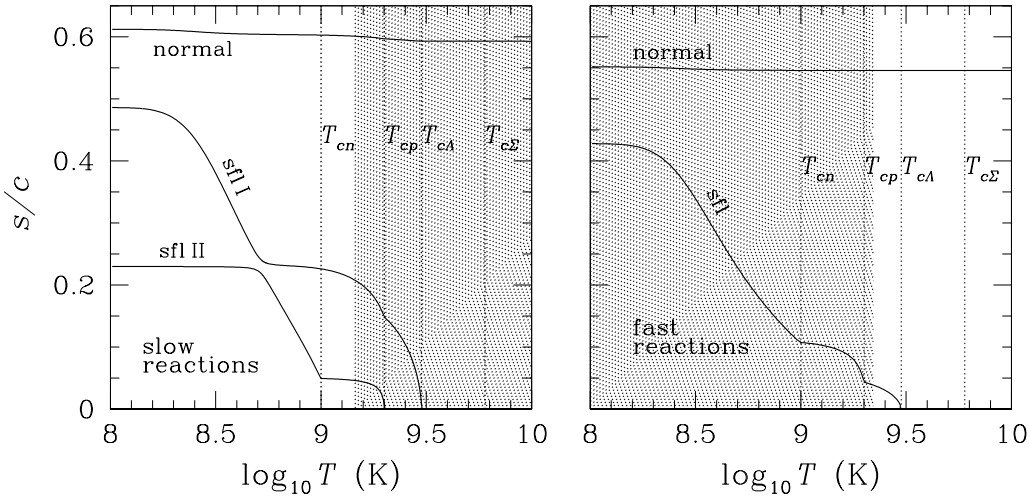}
\end{center}
\caption{
Speed of sound modes $s$ in units
of $c$ versus $T$
at $n_b=3 n_{b0}=0.48$ fm$^{-3}$
for the {\it third} equation of state 
of Glendenning \cite{glendenning85}.
Pulsation frequency is $\omega=10^4$ s$^{-1}$.
Left panel shows three sound modes
(`normal', `sfl I', and `sfl II')
in the limit of {\it slow} reactions.
Right panel shows two sound modes (`normal' and `sfl')
in the limit of {\it fast} reactions.
Baryon critical temperatures are indicated by vertical dots.
Range of $T$ where the limit
of {\it slow} (left panel) and {\it fast}
(right panel) reactions is invalid
(for $\omega=10^4$ s$^{-1}$), is shaded.}
\label{fig1}
\end{figure}
%
Figure \ref{fig1} illustrates
the results of numerical calculation
of the sound speeds $s$ (in units of $c$)
as functions of temperature $T$
in the limit of {\it slow} reactions (left panel)
and {\it fast} reactions (right panel).
Here and below in this paper we use
the {\it third} equation of state 
of Glendenning \cite{glendenning85}.
The figure is plotted for 
the baryon number density $n_b=3 n_0 =0.48$ fm$^{-3}$,
where $n_0=0.16$ fm$^{-3}$
is the nucleon number density in atomic nuclei.
%
The critical temperatures 
$T_{ci}$ ($i=n$, $p$, $\Lambda$, $\Sigma$) 
for transition of baryons 
to the superfluid state are poorly known.
We take them to be
$T_{cn}=10^9$ K,
$T_{cp}=2 \times 10^9$ K,
$T_{c\Lambda}=3 \times 10^9$ K, and
$T_{c\Sigma}= 6 \times 10^9$ K 
in accordance 
with some theoretical predictions 
(see, e.g., \cite{yls99,ls01,vt04,tnyt06}).
The relativistic entrainment matrix $Y_{ik}$
is taken from Ref.\ \cite{gkh08b}
(see also Ref.\ \cite{gkh08a},
where this matrix is calculated for $T=0$).

The shaded region in the left panel 
of Fig.\ \ref{fig1}
corresponds to temperatures, 
where the limit of {\it slow} reactions 
is invalid.
Rather conventionally, 
we define it by the inequality
$\lambda (\partial \delta \mu/\partial n_H
-\partial \delta \mu/\partial n_{\Sigma n}) > \omega/3$.
Similarly, 
the shaded region 
in the right panel of Fig.\ \ref{fig1}
indicates the range of temperatures,
where the limit of {\it fast} reactions is invalid.
We define this region by the inequality
$\lambda (\partial \delta \mu/\partial n_H
-\partial \delta \mu/\partial n_{\Sigma n}) < 3 \omega$.

Here and below we choose 
the frequency $\omega$ 
equal to $\omega=10^4$ s$^{-1}$.
In fact, this value of $\omega$
is more appropriate for studies 
of pulsating neutron stars
rather than sound waves,
because it results in wavelengths 
of the order of stellar radius. 
However, we believe that 
the relatively simple analysis
of sound waves 
with our choice of $\omega$
may provide some insight into the complex properties 
of global pulsations of neutron stars 
with superfluid nucleon-hyperon cores.

The rates of the reactions $\lambda_1$,$\ldots$,$\lambda_4$,
which constitute the quantity
$\lambda=\lambda_1+\lambda_2+\lambda_3+\lambda_4$,
are poorly known and depend essentially 
on the model of baryon-baryon interactions
and on the many-body theory employed
(see, e.g., \cite{hly02,lo02,dd04,jones01,schaffner08}).
It is natural to expect that the rates $\lambda_i$
are order-of-magnitude comparable \cite{jones01}.
In our numerical calculations
we, following Ref.\ \cite{lo02},
take into account only 
the contribution to $\lambda$ 
from the reactions (\ref{s}) and (\ref{l}) 
(i.e. we assume that $\lambda_3=\lambda_4=0$).
The rates $\lambda_{10}$ and $\lambda_{20}$
for {\it nonsuperfluid} matter 
are taken from Ref.\ \cite{lo02}.
The baryon superfluidity 
suppresses the reaction rates,
which can be presented as
$\lambda_1=\lambda_{10} R_1$ and $\lambda_2=\lambda_{20} R_2$.
The reduction factors $R_1 \leq 1$ and $R_2 \leq 1$
are calculated using the formula (28) of Ref.\ \cite{hly02}.
Though the authors of Ref.\ \cite{hly02}
proposed the formula for the reaction (\ref{s}),
it remains valid for the reaction (\ref{l}).
In both cases the index $i$ in that formula
enumerate the reacting particles
($i=n$, $n$, $p$, $\Sigma$ for the reaction (\ref{s}) and
$i=n$, $p$, $p$, $\Lambda$ for the reaction (\ref{l})).

As one can see from Fig.\ 1,
at $T<T_{cn}$ (when all baryon species are superfluid)
there are three sound modes in the limit of {\it slow} 
reactions and two in the limit of 
{\it fast} reactions.
With further increasing temperature 
the number of sound modes decreases.
As a result, at $T>T_{c \Lambda}$ 
there is only one mode in both limiting cases.
At $T>T_{c \Sigma}$ 
this mode becomes the ordinary sound in 
nonsuperfluid nucleon-hyperon matter
(in fact, it transforms into the ordinary sound
already at $T>T_{c \Lambda}$, see the next paragraph).
We term this sound mode normal;
in the figure it is denoted as `normal'.
Accordingly, the other modes are termed superfluid
and denoted as `sfl I', `sfl II' (see Fig.\ 1, left panel)
and `sfl' (see Fig.\ 1, right panel).
Notice that, the speed of normal mode
only weakly depends on $T$
both in the limits of {\it slow}
and {\it fast} reactions
and practically coincides 
with the speed of sound of normal matter.
On the contrary, the speeds of superfluid modes
strongly depend on temperature and approach their
asymptotic values only at $T \la 10^8$ K.

Let us explain,
how the number of sound modes changes with $T$
in the limit of {\it slow} reactions (Fig.\ 1, left panel).
Limit of {\it fast} reactions can be considered in a similar way.
As we have already discussed above, 
at $T<T_{cn}$ there are five velocities,
${\pmb u}$ and ${\pmb w}_{(i)}$ ($i=n$, $p$, $\Lambda$, $\Sigma$).
They are related by two conditions, 
(\ref{sound55g}) and (\ref{sound4g}).
Thus, the number of independent velocities
(and sound modes) equals three.
At $T_{cn}<T<T_{cp}$ neutrons are normal,
that is there are only four velocities in the system,
${\pmb u}$ and ${\pmb w}_{(i)}$ ($i=p$, $\Lambda$, $\Sigma$).
These velocities are constrained 
by the only one condition (\ref{sound55g})
(the condition (\ref{sound4g}) is not applicable; 
it is valid only if all baryon species are superfluid).
Thus, the number of sound modes remains equal three.
Then, at $T_{cp}<T<T_{c \Lambda}$
and at $T_{c \Lambda}<T<T_{c \Sigma}$
a motion with, respectively, 
three (${\pmb u}$, ${\pmb w}_{\Lambda}$,
and ${\pmb w}_{\Sigma}$)
and two (${\pmb u}$ and ${\pmb w}_{\Sigma}$) 
velocities is possible.
These velocities are related by the condition (\ref{sound55g}).
Consequently, there are two sound modes
in the range $T_{cp}<T<T_{c \Lambda}$
and one in the range $T_{c \Lambda}<T<T_{c \Sigma}$.
In the latter case it follows 
from the condition (\ref{sound55g})
that ${\pmb w}_{\Sigma}=0$.
Hence, the hydrodynamic equations 
are formally the same as 
those for normal liquid,
and the only sound mode coincides with 
the ordinary sound in nonsuperfluid matter.
Finally, at $T>T_{c \Sigma}$ all baryon species are normal
and move with the same velocity ${\pmb u}$.
Since this velocity is not constrained,
there is (as it should be) 
only one sound mode.

\section{Damping of sound waves}
\label{4}

\subsection{Damping times: general equations}

In the previous section we analysed the sound modes 
in the limit of {\it slow} and {\it fast} reactions 
neglecting dissipation.
In this section we allow for a weak dissipation 
in these limiting cases, 
which is due to the shear viscosity 
and nonequilibrium processes 
(\ref{s})--(\ref{ll}).
Our aim is the calculation of 
the characteristic damping times $\tau$ 
of sound waves (the so called e-folding times).

A few ways exist to calculate $\tau$.
For instance, 
one can explicitly take into account
the shear viscosity 
in Eqs.\  (\ref{sound55g})--(\ref{sound3g})
and the next (complex) terms in the expansion 
of $\delta n_H$ and $\delta n_{\Sigma n}$ 
into series in powers of 
$\lambda (\partial \delta \mu/\partial n_H
-\partial \delta \mu/\partial n_{\Sigma n})/\omega$
in the limit of {\it slow} reactions 
and in powers of 
$\omega/[\lambda (\partial \delta \mu/\partial n_H
-\partial \delta \mu/\partial n_{\Sigma n})]$
in the limit of {\it fast} reactions (see Appendix).
Then, solving the system of equations (\ref{delta P})--(\ref{sound3g}),
one can find the small complex correction $\delta s$ 
to the speed of sound $s$, 
which is related to the damping time $\tau$ by
\begin{eqnarray}
\tau=\frac{i}{k \delta s}.
\label{tau1}
\end{eqnarray}
Another way to calculate $\tau$ is to use
the effective bulk viscosity formalism \cite{gk08},
similar to how it was done in Ref.\ \cite{gusakov07}
for $npe$ matter.

We calculated $\tau$ by both methods
described above. 
However, here we present the 
most simple, third method of calculation.
Of course, all three methods give 
the same results.

Let us define the characteristic damping time as
\begin{eqnarray}
\tau \equiv - \frac{2E_{\rm puls}}
{\langle\dot{E}_{\rm puls}\rangle},
\label{tau}
\end{eqnarray}
where $E_{\rm puls}$ is the mechanical energy 
(per unit volume) of pulsations;
$\langle\dot{E}_{\rm puls}\rangle$ 
is the rate of change of $E_{\rm puls}$,
averaged over the period $2 \pi/\omega$. 
Here and below angle brackets denote averaging over 
the pulsation period.
The factor 2 in Eq.\ (\ref{tau}) appears because 
$\tau$ is the e-folding time 
for the hydrodynamic velocities
rather than for energy.
One can check that this definition of $\tau$
coincides with that given by Eq.\ (\ref{tau1}).

The mechanical energy of a pulsating superfluid matter
can be easily found if we notice that $E_{\rm puls}$ 
entirely transforms into its kinetic energy 
when the matter in the course of pulsations 
passes through the equilibrium point. 
Generally, the kinetic energy $E_{\rm kin}$ 
of the superfluid matter equals
(we remind that $u \ll c$ and $w_{(i)}/\mu_i \ll c$)
\begin{equation}
E_{\rm kin} = \frac{1}{2} \,
\left\{(P+\varepsilon) \, {\pmb u}^2+Y_{ik}[\mu_i \, {\pmb w}_{(k)}
{\pmb u}+\mu_k \, {\pmb w}_{(i)} {\pmb u}+{\pmb w}_{(i)}
{\pmb w}_{(k)}]\right\}.
\label{Ekin}
\end{equation}
In the nonrelativistic limit and for two-component 
mixture this formula agrees with the expression (7) 
of Ref.\ \cite{ab75}.
Since in a sound wave
the vectors ${\pmb u}$ and ${\pmb w}_{(i)}$
are collinear with ${\pmb k}$ and can be presented as
${\pmb u}={\pmb u}_a \cos(\omega t + {\pmb k} {\pmb r})$ 
and
${\pmb w}_{(i)}={\pmb w}_{(i)a} 
\cos(\omega t + {\pmb k} {\pmb r})$,
one obtains for $E_{\rm puls}$
($u_a \equiv \pmb u_a \pmb k/k$
and $w_{(i)a}\equiv \pmb w_{(i)a} \pmb k/k$)
\begin{eqnarray}
E_{\rm puls} = \frac{1}{2} \,
\left\{(P_0+\varepsilon_0) \, u_a^2+Y_{ik}[\mu_{i0} \, w_{(k)a}
u_a+\mu_{k0} \, w_{(i)a} u_a+w_{(i)a} w_{(k)a}]\right\}.
\label{Emeh}
\end{eqnarray}
Here we used the fact that at the equilibrium point 
the quantities ${\pmb u}$ and ${\pmb w}_{(i)}$
are equal to their amplitudes 
${\pmb u}_{a}$ and ${\pmb w}_{(i)a}$, respectively.

To find $\langle\dot{E}_{\rm puls}\rangle$
let us notice that all this energy goes into heat.
Thus, using Eqs.\ (\ref{entropy2}) and (\ref{dG}), one can write
(see also Refs.\ \cite{ll87,gyg05})
\begin{equation}
\langle\dot{E}_{\rm puls}\rangle
= -\langle T \partial_\mu \left(S u^{\mu} \right) \rangle
=-\langle \lambda (\delta \mu)^2\rangle +
\langle \tau^{\mu \nu}_{\rm sh} \,\, \partial_{\mu} u_{\nu} \rangle.
\label{Edot}
\end{equation}
It follows from this 
equation together with Eq.\ (\ref{tau})
that the characteristic damping time $\tau_{\rm bulk}$
due to the nonequilibrium reactions 
(\ref{s})--(\ref{ll}) equals
\begin{equation}
\tau_{\rm bulk} =\frac{2E_{\rm puls}}
{\langle \lambda (\delta \mu)^2\rangle},
\label{taubulk}
\end{equation}
while the characteristic damping time $\tau_{\rm sh}$
due to the shear viscosity is
\begin{equation}
\tau_{\rm sh} =-\frac{2E_{\rm puls}}
{\langle \tau^{\mu \nu}_{\rm sh} \,\,
\partial_{\mu} u_{\nu} \rangle}.
\label{taush}
\end{equation}

As a consequence of Eq.\ (\ref{dmu2}) (see Appendix),
in the limit of {\it slow} reactions
\begin{equation}
\langle \lambda (\delta \mu)^2\rangle =
\frac{\lambda}{2s^2}\left(
\frac{\partial\delta\mu}{\partial n_{b}} {J}_{b\;a}
+ \frac{\partial\delta\mu}{\partial n_{H}} {J}_{H\;a}
+ \frac{\partial\delta\mu}{\partial n_{\Sigma n}} {J}_{\Sigma n\;a}
\right)^2,
\label{Edot2s}
\end{equation}
while in the limit of {\it fast} reactions
\begin{equation}
\langle \lambda (\delta \mu)^2\rangle=
\frac{\omega^2}{2s^2 \lambda \left(\partial\delta\mu/\partial n_{H}
-\partial\delta\mu/\partial n_{\Sigma n}
\right)^2}\left(\frac{\partial\delta\mu}{\partial n_{b}}{J}_{b\;a}
+ \frac{\partial\delta\mu}{\partial n_{H}}{J}_{H\;a}
+ \frac{\partial\delta\mu}{\partial n_{\Sigma n}} {J}_{\Sigma n\;a}
\right)^2.
\label{Edot2f}
\end{equation}
In Eqs.\ (\ref{Edot2s}) and (\ref{Edot2f})
${J}_{b\;a}\equiv n_{b} {u_a}+\sum_i Y_{ik}{w}_{(k)a}$,
${J}_{H\;a}\equiv n_{H} {u_a}+Y_{\Sigma k}
{w}_{(k)a}+Y_{\Lambda k}{w}_{(k)a}$,
and ${J}_{\Sigma n\;a}\equiv n_{{\Sigma n}}{u_a}
+Y_{\Sigma k}{w}_{(k)a}+Y_{nk}{w}_{(k)a}$;
$s$ is the speed of sound calculated 
in the previous section neglecting dissipation;
the factor $1/2$ is a result of the averaging 
over the pulsation period.

The dissipation rate of the mechanical energy 
due to the shear viscosity is the same in both limits,
\begin{equation}
\langle \tau^{\mu \nu}_{\rm sh}
\,\, \partial_{\mu} u_{\nu} \rangle
= -\frac{2}{3} \, \eta \, \frac{\omega^2}{s^2} \, u_a^2.
\label{Edotshear}
\end{equation}
To obtain this formula we used Eq.\ (\ref{taush1}) 
for $\tau^{\mu \nu}_{\rm sh}$.
We see that $\langle \tau^{\mu \nu}_{\rm sh}
\,\, \partial_{\mu} u_{\nu} \rangle$ 
is formally given by the same expression 
as in the case of nonsuperfluid matter.

As follows from Eqs.\ (\ref{Emeh}), (\ref{Edot}), and 
(\ref{Edot2s})--(\ref{Edotshear}),
$E_{\rm puls}$ and $\langle\dot{E}_{\rm puls}\rangle$
are the functions of the amplitudes $u_a$ and $w_{(i)a}$.
Using the system of linear nondissipative equations
(\ref{delta P})--(\ref{sound3g}), 
the amplitudes $w_{(i)a}$ can be expressed through
$u_a$ and presented in the form
\begin{equation}
w_{(i)a}=\alpha_i(s) \, u_a,
\label{alpha}
\end{equation}
where $\alpha_i(s)$ are some coefficients depending on $s$; 
they differ for each sound mode.
In view of Eq.\ (\ref{alpha}), 
one gets from Eqs.\ (\ref{Emeh}), (\ref{Edot}), 
and (\ref{Edot2s})--(\ref{Edotshear}) 
that $E_{\rm puls} \sim u_a^2$
and $\langle\dot{E}_{\rm puls}\rangle \sim u_a^2$.
Hence, $\tau$ is independent of $u_a$ (see Eq.\ (\ref{tau})).

\subsection{Results for damping times}

We numerically calculate the coefficients $\alpha_i(s)$ 
from Eqs.\ (\ref{delta P})--(\ref{sound3g}) 
and thus determine the characteristic damping times $\tau$
of sound modes in the limit of {\it slow} and {\it fast} reactions.
Figures 2, 3, 4, and 5 illustrate the results of our calculations.
These figures are plotted assuming the same microphysics input 
(the baryon number density, the critical baryon temperatures etc)
as in Fig.\ 1.

\begin{figure}
\begin{center}
\leavevmode \epsfxsize=14cm \epsfbox[30 310 480 520]{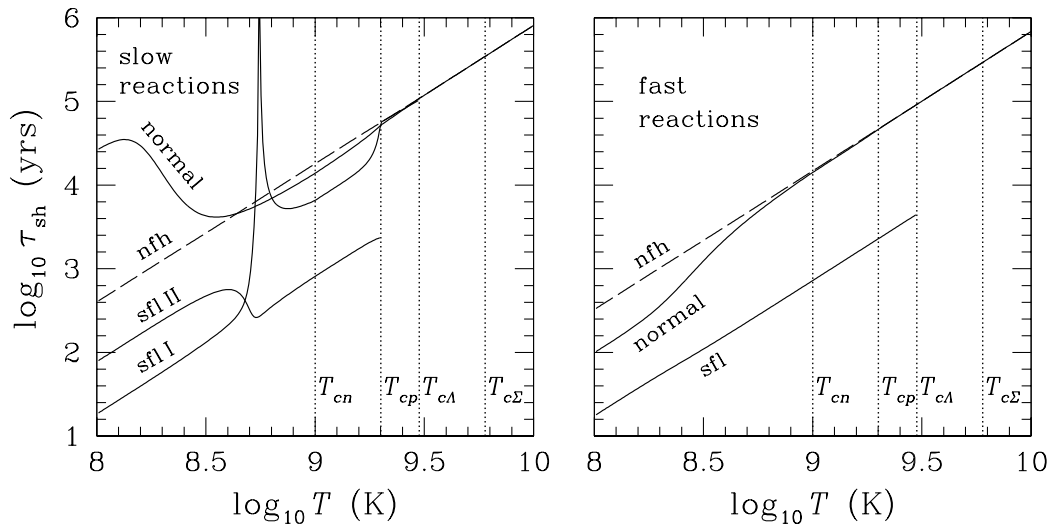}
\end{center}
\caption{
The characteristic damping times $\tau_{\rm sh}$
due to the shear viscosity versus $T$ 
in the limit of {\it slow} reactions (left panel)
and {\it fast} reactions (right panel).
Dashed curves (marked with `nfh') in both panels
are obtained using the nonsuperfluid hydrodynamics, 
see the text. 
Other notations are the same as in Fig.\ 1.
}
\label{fig2}
\end{figure}
Figure 2 shows the characteristic damping times $\tau_{\rm sh}$
due to the shear viscosity as a function of  temperature $T$
in the limit of {\it slow} reactions (left panel) 
and {\it fast} reactions (right panel).
To calculate $\tau_{\rm sh}$ it is necessary to know
the shear viscosity coefficient $\eta$ of superfluid 
nucleon-hyperon matter.
This coefficient has not been considered in the literature so far.
For definiteness, we take for $\eta$
the shear viscosity of electrons and muons 
$\eta_{e \mu}=\eta_{e}+\eta_{\mu}$
from Ref.\ \cite{sy08}.
In this reference it is shown that
(for $npe\mu$ matter) 
the contribution of $\eta_{e \mu}$ to the
total shear viscosity $\eta$ is dominant.
Notice, however, that this result 
was obtained under assumption 
that only protons are possibly superfluid
(neutrons were treated as normal).
When plotting Fig.\ 2 we used 
the coefficient $\eta_{e \mu}$
calculated for {\it nonsuperfluid} matter 
from Eq.\ (37) of Ref.\ \cite{sy08}.
The effects of baryon superfluidity on $\tau_{\rm sh}$
are illustrated in the next figure.

By the solid curves we show $\tau_{\rm sh}$ calculated 
for each sound mode
by means of Eq.\ (\ref{taush}).
Dashes demonstrate the characteristic 
damping times $\tau_{\rm nfh-sh}$
calculated using the simplified model, 
the hydrodynamics of normal (nonsuperfluid) liquid.
In this case there is only one mode in both limits;
the corresponding curves are marked with `nfh', 
which is the abbreviation 
of `normal fluid hydrodynamics'.
It is straightforward to demonstrate that
$\tau_{\rm nfh-sh}=3(P_0+\varepsilon_0) s^2/(2\omega^2\eta)$ 
(see, e.g., Ref.\ \cite{ll87}).
As it should be, 
$\tau_{\rm sh}$ for normal mode 
(in the figure it is marked `normal') 
coincides with $\tau_{\rm nfh-sh}$
at $T>T_{c\Lambda}$ 
(see Sec.\ IVC).

It follows from Fig.\ 2 that $\tau_{\rm sh}$, 
calculated in the frame of relativistic hydrodynamics
of superfluid mixtures, 
can strongly (by several orders of magnitude)
differ from $\tau_{\rm nfh-sh}$.
It is interesting that the maximum deviation of 
$\tau_{\rm sh}$ from $\tau_{\rm nfh-sh}$ 
at low temperatures ($T \la 3 \times 10^8$ K)
is observed for {\it normal} mode (Fig.\ 2, left panel),
though it is analogous
to the usual sound in nonsuperfluid matter.

At temperature $T \approx 5.55\times10^8$ K
the characteristic damping time $\tau_{\rm sh}$ for
one of the superfluid modes (`sfl I') becomes infinite.
This is because at such temperature and for this mode
the hydrodynamic motions occur in such a way 
that the normal component is always at rest ($u_a=0$), 
while the superfluid components pulsate around it.
Mathematically, this means that the coefficients $\alpha_i(s)$
in Eq.\ (\ref{alpha}) are infinite.
It follows then from Eqs.\ (\ref{Emeh}), (\ref{taush}), 
and (\ref{Edotshear}) 
that dissipation 
due to the shear viscosity is absent. 

\begin{figure}
\begin{center}
\leavevmode \epsfxsize=16cm \epsfbox[15 310 564 482]{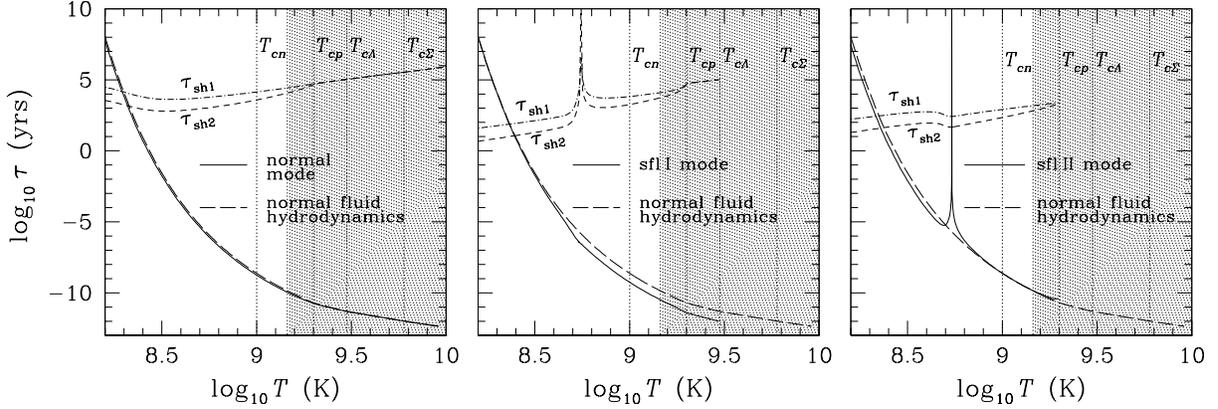}
\end{center}
\caption{
The characteristic damping times $\tau$ versus $T$
for `normal' mode (left panel),
`sfl I' mode (middle panel),
and `sfl II' mode (right panel)
in the limit of {\it slow} reactions.
Solid curves demonstrate
$\tau_{\rm bulk}$
calculated from Eq.\ (\ref{taubulk});
long-dashed curve (the same in all panels)
describes the characteristic damping times 
$\tau_{\rm nfh-bulk}$,
calculated using the normal fluid hydrodynamics.
Dot-dashed and short-dashed curves
show damping times $\tau_{\rm sh1}$
and $\tau_{\rm sh2}$
due to the shear viscosity (see the text for more details).
Other notations are the same as in Figs.\ 1 and 2.}
\label{fig3}
\end{figure}
%
Figure 3 presents the dependence of the characteristic 
damping times on $T$ in the limit 
of {\it slow} reactions.
Three panels correspond to three modes 
(from left to right: `normal', `sfl I', and `sfl II').
The damping times $\tau_{\rm bulk}$,
calculated for each mode using Eq.\ (\ref{taubulk}),
are shown by solid curves.
For comparison, by long dashes we show 
the characteristic damping times $\tau_{\rm nfh-bulk}$
due to the nonequilibrium reactions (\ref{s})--(\ref{ll}),
which are obtained using 
the hydrodynamics of {\it nonsuperfluid} liquid.
When plotting the long-dashed curve 
superfluidity of baryons was taken into account 
only at calculating 
the total reaction rate $\lambda$.
Because there is only one sound mode 
in the nonsuperfluid hydrodynamics, 
this curve is the same in all three panels.

One can see from the figure that for the normal mode
the solid and long-dashed curves practically 
(on the logarithmic scale) coincide; 
they differ by a factor of 2 or less.
On the contrary, $\tau_{\rm bulk}$ for superfluid modes
can differ from $\tau_{\rm nfh-bulk}$
by orders of magnitude.

Notice that, at $T \approx 5.41 \times 10^8$ K
the damping time for the superfluid mode `sfl II' 
becomes infinite.
The point is at such temperature and for this mode the condition
$\delta \mu=0$ is always preserved during the pulsations.
The denominator in Eq.\ (\ref{taubulk}) is then vanished
and $\tau_{\rm bulk}$ tends to infinity.

\begin{figure}
\begin{center}
\vskip 5mm
\leavevmode \epsfxsize=8cm \epsfbox[140 300 400 500]{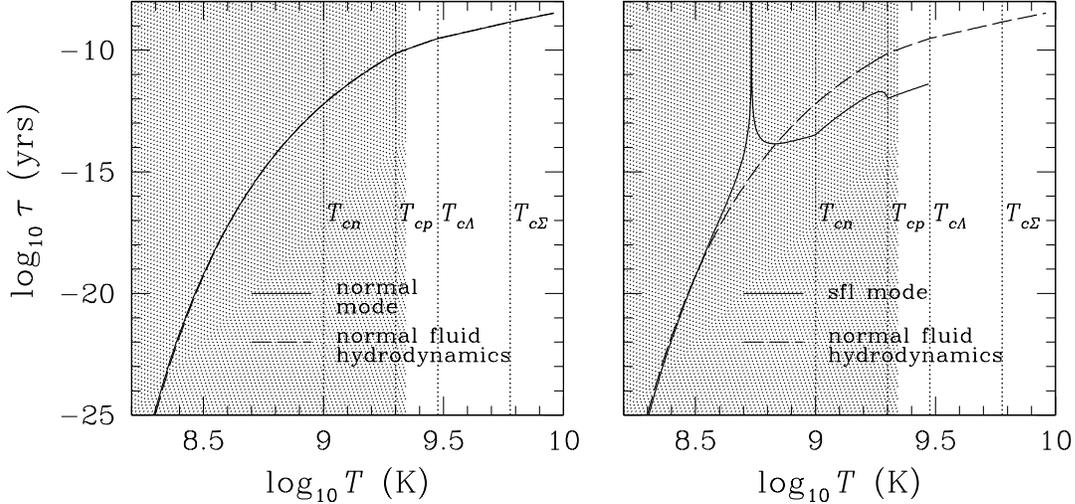}
\end{center}
\caption{
The same as in Fig.\ 3 but in the limit of {\it fast} reactions.
The damping times $\tau_{\rm sh} \gg \tau_{\rm bulk}$
and are not shown.
}
\label{fig4}
\end{figure}

%
By dot-dashes and short dashes in Fig.\ 3
we show, respectively, 
the characteristic damping times 
$\tau_{\rm sh1}$ and $\tau_{\rm sh2}$ 
due to the shear viscosity.
The times $\tau_{\rm sh1}$ 
are the same as in Fig.\ 2 (left panel);
they are plotted for the shear viscosity $\eta=\eta_{e \mu}$
of nonsuperfluid matter 
(see equation (37) of Ref.\ \cite{sy08}).
To obtain the dependence $\tau_{\rm sh2}(T)$ 
we assume that $\eta=\eta_{e \mu}$ as before,
but additionally take into account 
the reduction of $\eta_{e \mu}$ 
by proton superfluidity.
The reduction factor was calculated from
Eq.\ (83) of Ref.\ \cite{sy08}.
It is worth noting that this formula 
is obtained for nucleon $npe\mu$ matter 
and does not 
imply the superfluidity 
of other baryon species except for protons.
Thus, the dependence $\tau_{\rm sh2}(T)$ 
only qualitatively describes possible effect 
of baryon superfluidity on $\tau_{\rm sh}$.

It follows from the analysis of Fig.\ 3, 
that at high enough 
temperatures $T \ga 3 \times 10^8$ K,
the dissipation due to the shear viscosity 
is negligible in comparison 
to that due to the nonequilibrium processes 
(\ref{s})--(\ref{ll}). 
Moreover, the threshold density, 
at which $\tau_{\rm bulk} \approx \tau_{\rm sh}$,
only weakly depends on $\eta$.

\begin{figure}
\begin{center}
\leavevmode \epsfxsize=8cm \epsfbox[60 200 560 670]{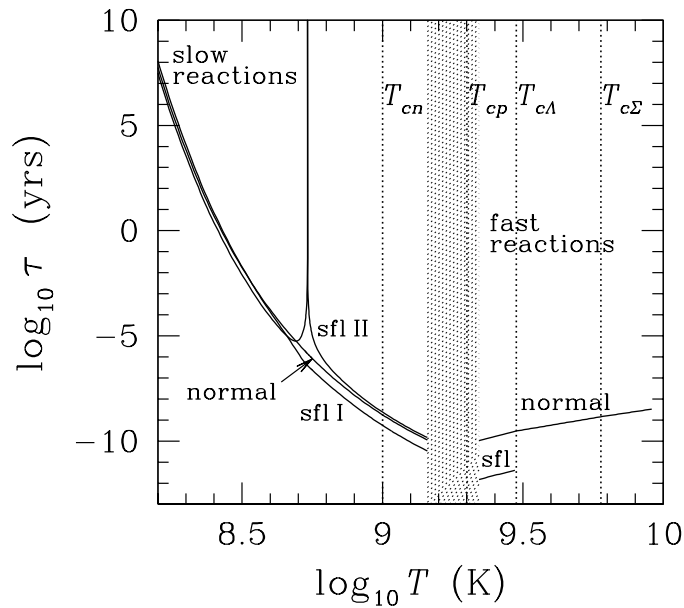}
\end{center}
\caption{
Characteristic damping times $\tau_{\rm bulk}$
versus $T$ in the limit of {\it slow}
($T \la 1.4 \times 10^{9}$ K)
and {\it fast} ($T \ga 2.3 \times 10^9$ K) reactions.
The region of $T$ where the rate of nonequilibrium reactions
is intermediate (i.e., where both limits are invalid),
is shaded.
}
\label{fig5}
\end{figure}

%
Figure 4 is analogous to Fig.\ 3
but is plotted for the limit of {\it fast} reactions.
The solid curve in the left panel demonstrates 
the dependence $\tau_{\rm bulk}(T)$ 
for the normal mode (`normal') and
in the right panel -- for the superfluid mode (`sfl').
The long-dashed curves are the same in both panels 
and describe the damping times $\tau_{\rm nfh-bulk}(T)$.
In the limit of {\it fast} reactions 
the damping times due to the shear viscosity 
are much greater than $\tau_{\rm bulk}$
and are not shown.

As in the limit of {\it slow} reactions,
$\tau_{\rm bulk}$ for the normal mode does not differ
substantially from $\tau_{\rm nfh-bulk}$
(no more than by a factor of 2).
However, at high enough temperatures
$\tau_{\rm bulk}$ for superfluid mode
is two orders of magnitude smaller 
than $\tau_{\rm nfh-bulk}$.

The analysis of Fig.\ 4 shows that
the damping time for superfluid mode becomes infinite 
at the same temperature $T\approx 5.41\times10^8$ K
as in the limit of {\it slow} reactions
(the corresponding peak is in the shaded region where
the limit of {\it fast} reactions is not applicable 
for the chosen pulsation frequency $\omega=10^4$ s$^{-1}$).
It is easy to understand why this peak fall on the same
temperature.
As we already discussed above, 
the infinite damping time $\tau_{\rm bulk}$ means that 
the condition $\delta \mu=0$ 
holds in the pulsating matter (at given temperature).
In this case, the nondissipative 
Eqs.\ (\ref{sound55g})--(\ref{sound3g})
in the limits of {\it slow} and {\it fast} 
reactions coincide.
Thus, the functional dependence of the amplitudes $w_{(i)a}$ 
on $u_a$ is the same in both limits (see Sec.\ VA). 
Consequently, 
as follows from 
Eqs.\ (\ref{Edot2s}) and (\ref{Edot2f}),
if $\delta \mu=0$ in one limit, 
then it vanishes in the other limit.

Fig.\ 5 is a combined plot, 
showing the characteristic damping times $\tau_{\rm bulk}$
as functions of $T$ 
for the limits of {\it slow} and {\it fast} reactions.
The shaded region (at $T \sim 2 \times 10^9$ K)
corresponds to intermediate temperatures 
at which both limits are inapplicable.
In this region the pulsation energy dissipates
on a time scale of order of the pulsation period.
Notice that, at $T \ga 3\times10^8$~K 
the characteristic damping time of pulsations 
$\tau=\tau_{\rm bulk}+\tau_{\rm sh}$
practically coincides with $\tau_{\rm bulk}$,
since at such temperatures 
the contribution of the shear viscosity to dissipation
is negligible.

\section{Summary}
\label{5}

In Ref.\ \cite{gk08} 
the relativistic dissipative hydrodynamics
was suggested to describe 
superfluid nucleon-hyperon matter 
in the cores of massive neutron stars.
Using this hydrodynamics, 
we analyse the sound waves, 
which are the simplest example of pulsations 
in such matter.

We demonstrate that in the limit of {\it slow} 
reactions (\ref{s})--(\ref{ll})
(when the composition of pulsating matter
is practically unaffected by these reactions)
there are {\it three} sound modes: one normal 
and two superfluid. 
In the opposite limit of {\it fast} reactions 
(when the pulsating matter is nearly at equilibrium 
with respect to the reactions (\ref{s})--(\ref{ll})),
only {\it two} sound modes exist:
the normal one 
and the superfluid one. 
In the intermediate case the sound waves cannot propagate
because they are damped on a time scale 
of order of the pulsation period.

The speed of normal 
sound mode 
in both limits is practically independent of
temperature and coincides with the sound speed 
for nonsuperfluid matter.
This mode turns into the ordinary sound 
at $T>T_{c \Lambda}$. 
On the contrary, the speeds of superfluid modes
strongly depend on temperature 
and vanish before the transition of matter 
to the normal state.

We analyse also the characteristic damping times $\tau$
of sound modes (Figs.\ 2, 3, 4, and 5).
We allow for the two main dissipative mechanisms:
damping 
due to the shear viscosity and
due to the nonequilibrium reactions (\ref{s})--(\ref{ll}) 
(these reactions generate the effective bulk viscosity).
We demonstrate that
($i$) the damping times $\tau$ 
for normal and superfluid modes 
can differ from each other by orders of magnitude;
($ii$) at $T \ga 3 \times 10^8$ K 
the damping due to the nonequilibrium 
reactions (\ref{s})--(\ref{ll}) 
is the dominant mechanism 
of dissipation;
this result is nearly insensitive 
to an actual value of the shear viscosity 
coefficient $\eta$.

In addition, we compare $\tau$
with the damping time $\tau_{\rm nfh}$,
calculated using the ordinary nonsuperfluid hydrodynamics,
but taking into account the effects of superfluidity 
on the shear viscosity and 
on the rates of the reactions 
(\ref{s})--(\ref{ll}). 
We show that ($iii$) $\tau$ approximately 
(up to a factor of two) 
coincides with $\tau_{\rm nfh}$ 
only for the normal mode and 
under the condition that 
the shear viscosity can be neglected
(i.e. $T \ga 3 \times 10^8$ K).
In other cases 
(for the superfluid modes 
and for the normal mode at $T < 3 \times 10^8$ K)
$\tau$ can differ from $\tau_{\rm nfh}$ by
several orders of magnitude.

The results listed above are obtained 
from the analysis of sound waves in 
the superfluid nucleon-hyperon matter.
However, they can serve as an indication 
that the effects, related to difference 
between the superfluid and normal fluid hydrodynamics,
can also be very important in studies 
of global pulsations of superfluid neutron stars,
essentially modifying their damping times.

\section*{Appendix}

Let us calculate the quantities $\delta n_b = n_b-n_{b0}$,
$\delta n_H=n_H-n_{H0}$,
$\delta n_{\Sigma n}=n_{\Sigma n}-n_{\Sigma n0}$,
and $\delta y=y-y_0$,
entering Eqs.\ (\ref{delta P})--(\ref{dmu})
for $\delta P$, $\delta \mu_{n}$,
and $\delta \mu$, respectively.
For that, we make use of the continuity equations
(\ref{particle_conservation2}),
assuming the perturbations are harmonic.
In the linear approximation
the continuity equations for leptons (electrons and muons)
have the form ($l=e$, $\mu$)
\begin{equation}
i \omega \, \delta n_{l}
+i {\pmb k} \, n_{l0} {\pmb u}=0,
\label{leptons_eq}
\end{equation}
where we put $\Delta S_l=0$, 
because
the leptonic reactions are slow.
From these equations it follows that
\begin{equation}
\delta y=0.
\label{y}
\end{equation}
%
The continuity equations for baryons, hyperons,
and $\Sigma^-$-hyperons with neutrons
can also be obtained from Eq.\ (\ref{particle_conservation2})
\begin{eqnarray}
i \omega \, \delta n_{b}
+i {\pmb k}{\pmb J}_b &=& 0,
\label{contin b}\\
i \omega \, \delta n_{H}
+i {\pmb k}
{\pmb J}_H &=& \lambda \delta \mu,
\label{contin H} \\
i \omega \, \delta n_{\Sigma n}
+i {\pmb k}
{\pmb J}_{\Sigma n} &=& -\lambda \delta \mu.
\label{contin Sn}
\end{eqnarray}
Here we used Eq.\ (\ref{dG}),
and introduced the notations
$\pmb{J}_{b}\equiv n_{b} \pmb{u}+\sum_i Y_{ik}\pmb{w}_{(k)}$;
$\pmb{J}_{H}\equiv n_{H} \pmb{u}+Y_{\Sigma k}
\pmb{w}_{(k)}+Y_{\Lambda k}\pmb{w}_{(k)}$; and
$\pmb{J}_{\Sigma n}\equiv n_{{\Sigma n}}
\pmb{u} +Y_{\Sigma k}\pmb{w}_{(k)}+Y_{nk}\pmb{w}_{(k)}$.
Solving now the system
(\ref{contin b})-(\ref{contin Sn})
taking into account Eqs.\ (\ref{dmu}) and (\ref{y}),
one gets for $\delta n_{b}$,
$\delta n_{H}$, and $\delta n_{\Sigma n}$
\begin{eqnarray}
\delta n_b &=&-\frac{{\pmb k}{\pmb J}_b}{\omega},
\label{dnb}\\
\delta n_H &=& \frac{{\pmb k}}{\omega}
\; \frac{i \omega {\pmb J}_H  + \lambda
[{\pmb J}_b\; \partial \delta \mu/\partial n_b+({\pmb J}_H
+{\pmb J}_{\Sigma n}) \; \partial \delta \mu
/\partial n_{\Sigma n}]}{\lambda (\partial \delta \mu/\partial n_H
-\partial \delta \mu/\partial n_{\Sigma n})-i \omega},
\label{dnH}\\
\delta n_{\Sigma n} &=&
\frac{{\pmb k}}{\omega} \; \frac{i \omega {\pmb J}_{\Sigma n}
- \lambda [{\pmb J}_b\;
\partial \delta \mu/\partial n_b+({\pmb J}_H+{\pmb J}_{\Sigma n})
\; \partial \delta \mu/\partial n_{H}]}{\lambda
(\partial \delta \mu/\partial n_H
-\partial \delta \mu/\partial n_{\Sigma n})-i \omega}.
\label{dnSn}
\end{eqnarray}
Using these equalities,
Eq.\ (\ref{dmu}) for $\delta \mu$
can be rewritten as
\begin{eqnarray}
\delta \mu =-
\frac{{\pmb k}}{i \lambda \; \left(\partial\delta\mu/\partial n_{H} -
\partial\delta\mu/\partial n_{\Sigma n} \right)
+ \omega}\left(\frac{\partial\delta\mu}{\partial n_{b}}
\pmb{J}_{b} + \frac{\partial\delta\mu}{\partial n_{H}}
\pmb{J}_{H} + \frac{\partial\delta\mu}
{\partial n_{\Sigma n}}  \pmb{J}_{\Sigma n} \right).
\label{dmu2}
\end{eqnarray}

In the limit of {\it slow} reactions,
when the total rate $\lambda$ of the reactions 
(\ref{s})--(\ref{ll}) is small,
that is
$\lambda (\partial \delta \mu/\partial n_H
-\partial \delta \mu/\partial n_{\Sigma n})\ll \omega$, 
one has
\begin{eqnarray}
\delta n_b=-{\pmb k}{\pmb J}_b/\omega,
\label{dnb slow}\\
\delta n_H=-{\pmb k}{\pmb J}_H/\omega,
\label{dnH slow}\\
\delta n_{\Sigma n}=-{\pmb k}{\pmb J}_{\Sigma n}/\omega.
\label{dnSn slow}
\end{eqnarray}
In the limit of {\it fast} reactions,
when $\lambda (\partial \delta \mu/\partial n_H
-\partial \delta \mu/\partial n_{\Sigma n})\gg \omega$,
one obtains
\begin{eqnarray}
\delta n_b &=& -\frac{{\pmb k}{\pmb J}_b}{\omega},
\label{dnb fast}\\
\delta n_H &=& \frac{{\pmb k}({\pmb J}_b\;
\partial \delta \mu/\partial n_b
+({\pmb J}_H+{\pmb J}_{\Sigma n}) \;
\partial \delta \mu/\partial n_{\Sigma n})}
{\omega(\partial \delta \mu/\partial n_H-\partial
\delta \mu/\partial n_{\Sigma n})},
\label{dnH fast}\\
\delta n_{\Sigma n} &=&-\frac{{\pmb k}({\pmb J}_b
\;\partial \delta \mu/\partial
n_b+({\pmb J}_H+{\pmb J}_{\Sigma n})\;
\partial \delta \mu/\partial
n_{H})}{\omega(\partial \delta \mu/\partial n_H
-\partial \delta \mu/\partial n_{\Sigma n})}.
\label{dnSn fast}
\end{eqnarray}
One sees, that in the both limits $\delta n_b$,
$\delta n_H$, and $\delta n_{\Sigma n}$ 
are real-valued and 
do not depend on the total reaction rate $\lambda$.
Correspondingly, 
the sound speeds are also real-valued, 
or, in other words, the dissipation is absent.
The dissipation due to the weak nonequilibrium processes
(\ref{s})--(\ref{ll}) can be taken into account 
by considering the next (complex) terms 
in the expansion of $\delta n_H$ and $\delta n_{\Sigma n}$
into series in powers of 
$\lambda (\partial \delta \mu/\partial n_H
-\partial \delta \mu/\partial n_{\Sigma n})/\omega$
in the case of {\it slow} reactions
and in powers of 
$\omega/[\lambda (\partial \delta \mu/\partial n_H
-\partial \delta \mu/\partial n_{\Sigma n})]$
in the case of {\it fast} reactions.

\section*{Acknowledgments}

The authors are very grateful to
K.P. Levenfish and D.G. Yakovlev for allowing
to use their code which calculates the third equation
of state of Glendenning \cite{glendenning85}.
This research was supported in part
by RFBR (Grants 08-02-00837 and 05-02-22003)
and by the Federal Agency for Science and Innovations
(Grant NSh 2600.2008.2).
One of the authors (M.E.G.) also acknowledges
support from the Dynasty Foundation
and from the RF Presidential Program
(grant MK-1326.2008.2).



\begin{thebibliography}{999}

\bibitem{sw2005}
T. E. Strohmayer and A. L. Watts,
Astrophys. J. \textbf{632}, 111 (2005).

\bibitem{israel2005}
G. L. Israel, T. Belloni, L. Stella,
Y. Rephaeli, D. E. Gruber, P. Casella,
S. Dall'Osso, N. Rea, M. Persic,
and R. E. Rothschild,
Astrophys. J. \textbf{628}, L53 (2005).

\bibitem{ak01}
N. Andersson and K. D. Kokkotas,
Int. J. Mod. Phys. \textbf{D10}, 381 (2001).


\bibitem{andersson03}
N. Andersson,
Class. Quantum Grav.
\textbf{20}, R105 (2003).


\bibitem{andersson06}
N. Andersson,
Astrophys. Space Sci. \textbf{308}, 395 (2007).

\bibitem{ligo07}
B. Abbott, R. Abbott, R. Adhikari,
J. Agresti, P. Ajith, B. Allen,
R. Amin, S. B. Anderson, W. G. Anderson,
M. Arain, and 437 coauthors,
Phys. Rev. \textbf{D76}, 062003 (2007).

\bibitem{yls99}
D. G. Yakovlev, K. P. Levenfish, and Yu. A. Shibanov,
Phys. Usp. {\bf 42}, 737 (1999).


\bibitem{ls01}
U. Lombardo and H.-J. Schulze,
Lect. Notes Phys. \textbf{578}, 30 (2001).


\bibitem{yp04}
D. G. Yakovlev and C. J. Pethick,
Annu. Rev. Astron. Astrophys. {\bf 42}, 169 (2004).

\bibitem{bb98}
Sh. Balberg and N. Barnea,
Phys. Rev. \textbf{C57}, 409 (1998).

\bibitem{vt04}
I. Vida${\tilde {\rm n}}$a and L. Tolos,
Phys. Rev. {\bf C70}, 028802 (2004).

\bibitem{tnyt06}
T. Takatsuka, S. Nishizaki, 
Y. Yamamoto, and R. Tamagaki,
Prog. Theor. Phys. {\bf 115}, 355 (2006).


\bibitem{epstein88}
R. I. Epstein,
Astrophys. J. \textbf{333}, 880 (1988).

\bibitem{ac01}
N. Andersson and G. L. Comer,
Mon. Not. R. Astron. Soc. \textbf{328}, 1129 (2001).

\bibitem{ga06}
M. E. Gusakov and N. Andersson,
Mon. Not. R. Astron. Soc. \textbf{372}, 1776 (2006).

\bibitem{gusakov07}
M.E. Gusakov, Phys. Rev. \textbf{D76}, 083001 (2007).


\bibitem{lm94}
L. Lindblom and G. Mendell,
Astrophys. J. \textbf{421}, 689 (1994).

\bibitem{lee95}
U. Lee,
Astron. Astrophys. \textbf{303}, 515 (1995).

\bibitem{ac01a}
L. Lindblom and G. Mendell,
Phys. Rev. \textbf{D61}, 104003 (2000).

\bibitem{acl02}
N. Andersson, G. L. Comer, and D. Langlois,
Phys. Rev. \textbf{D66}, 104002 (2002).

\bibitem{pca04}
R. Prix, G. L. Comer, and N. Andersson,
Mon. Not. R. Astron. Soc. \textbf{348}, 625 (2004).

\bibitem{yl03a}	
S. Yoshida and U. Lee,
Mon. Not. R. Astron. Soc. \textbf{344}, 207 (2003).

\bibitem{yl03b}
S. Yoshida and U. Lee,
Phys. Rev. \textbf{D67}, 124019 (2003).

\bibitem{cll99}
G. L. Comer, D. Langlois, and L. M. Lin,
Phys. Rev. \textbf{D60}, 104025 (1999).

\bibitem{sac08}
T. Sidery, N. Andersson, and G. L. Comer,
Mon. Not. R. Astron. Soc. \textbf{385}, 335 (2008).

\bibitem{lac08}
L.-M. Lin, N. Andersson, and G. L. Comer,
Phys. Rev. \textbf{D78}, 083008 (2008).


\bibitem{gk08}
M. E. Gusakov and E. M. Kantor,
Phys. Rev. {\textbf{D78}}, 083006 (2008).

\bibitem{jones01}
P. B. Jones,
Phys. Rev. \textbf{D64}, 084003 (2001).

\bibitem{hly02}
P. Haensel, K. P. Levenfish, and D. G. Yakovlev,
Astron. Astrophys. \textbf{381}, 1080 (2002).


\bibitem{lo02}
L. Lindblom and B. J. Owen,
Phys. Rev. \textbf{D65}, 063006 (2002).


\bibitem{schaffner08}
J. Schaffner-Bielich,
arXiv:0801.3791v1 (2008).

\bibitem{gkh08a}
M. E. Gusakov, E. M. Kantor, and P. Haensel,
Phys. Rev. {\textbf C}, submitted (2008).

\bibitem{gkh08b}
M. E. Gusakov, E. M. Kantor, and P. Haensel,
in preparation,
to be submitted to Phys. Rev.~{\textbf C}.


\bibitem{ab75}
A. F. Andreev and E. P. Bashkin,
Zh. Eksp. Teor. Fiz., \textbf{69}, 319 (1975).


\bibitem{bjk96}
M. Borumand, R. Joynt, and W. Klu$\acute{\rm z}$niak,
Phys. Rev. \textbf{C54}, 2745 (1996).


\bibitem{gh05}
M. E. Gusakov and P. Haensel,
Nucl. Phys. \textbf{A761}, 333 (2005).


\bibitem{khalatnikov89}
I. M. Khalatnikov,
{\it An Introduction to the Theory of Superfluidity}
(Addison-Wesley, New York, 1989).


\bibitem{putterman74}
S. J. Putterman,
{\it Superfluid Hydrodynamics}
(North-Holland, Amsterdam, 1974).



\bibitem{gyg05}
M. E. Gusakov, D. G. Yakovlev, and O. Y. Gnedin,
Mon. Not. R. Astron. Soc. \textbf{361}, 1415 (2005).



\bibitem{reisenegger95}
A. Reisenegger,
Astrophys. J. \textbf{442}, 749 (1995).


\bibitem{glendenning85}
N. Glendenning,
Astrophys. J. \textbf{293}, 470 (1985).


\bibitem{dd04}
E. N. E. van Dalen
and A. E. L. Dieperink,
Phys. Rev. \textbf{C69}, 025802 (2004).


\bibitem{ll87}
L. D. Landau and E. M. Lifshitz,
{\it Fluid mechanics},
Course of theoretical physics,
(Pergamon Press, Oxford, 1987).


\bibitem{sy08}
P. S. Shternin and D. G. Yakovlev,
Phys. Rev. \textbf{D78}, 063006 (2008).


\end{thebibliography}
\end{document}